\documentclass[useAMS,usenatbib]{mn2e}
\usepackage{graphicx,psfig,cite,epsf}  
\usepackage{times}
\usepackage{subfigure}
\usepackage{textcomp}
\usepackage{multirow}
\usepackage{amsmath}
\usepackage{amsmath}	% Advanced maths commands
\usepackage{amssymb}	% Extra maths symbols
\usepackage{gensymb}

\def\apjl{ApJ}

\def\apjl{ApJ}                % Astrophysical Journal, Letters
\def\apjs{ApJS}               % Astrophysical Journal, Supplement
\def\ao{Appl.~Opt.}           % Applied Optics
\def\aap{A\&A}                % Astronomy and Astrophysics
\def\src {AX J1845--0258}
\def\srcasca {AX J184453--025640}

\def\srccc {CXOU J184507.2--025657}
\def\srcccc {CXOU J184509.8--025714}

\title[The possible counterparts of AX J1845-0258]{A new investigation of the possible X-ray counterparts of the magnetar candidate AX J1845-0258}

\author[Pintore F.]{Fabio  Pintore$^{1}$\thanks{E-mail: pintore@iasf-milano.inaf.it}
and Sandro Mereghetti$^{1}$ \\
$^{1}$INAF - IASF Milano, Via E. Bassini 15, I-20133 Milano, Italy\\
}

\date{Accepted XXX. Received YYY; in original form ZZZ}

\pubyear{2015}

\begin{document}
\label{firstpage}
\pagerange{\pageref{firstpage}--\pageref{lastpage}}
\maketitle

\begin{abstract}

\src\ is a transient X-ray pulsar, with spin period of 6.97s,  discovered with the  {\it ASCA} satellite in 1993. Its soft spectrum and the possible association with a supernova remnant suggest that \src\ might be a magnetar, but this has not been confirmed yet. A  possible  counterpart one order of magnitude fainter, \srcasca ,  has been found in later X-ray observations,  but  no pulsations have been  detected. In addition, some other X-ray sources are compatible with the pulsar location, which is in a crowded region of the Galactic plane.
We have carried out a new investigation of all the X-ray sources in the   {\it ASCA} error region of \src , using archival data obtained with
{\it Chandra}  in 2007 and 2010, and with  {\it XMM-Newton} in 2010.  We set an upper limit of 6\%  on the pulsed fraction of   \srcasca\ and confirmed its rather hard spectrum (power law photon index of 1.2$\pm$0.3). In addition to the other two fainter sources already reported in the literature, we found other X-ray sources positionally consistent with \src . Although many of them are possibly foreground stars likely unrelated to the pulsar, at least another new source, CXOU J184457.5--025823, could be a plausible counterpart of \src.  It has a flux of $6\times10^{-14}$ erg cm$^{-2}$ s$^{-1}$ and a spectrum well fit by a power law with photon index  $\sim1.3$ and N$_\text{H}\sim10^{22}$ cm$^{-2}$.

\end{abstract}


\section{Introduction}

\src\ is an X-ray pulsar, with spin period of 6.97s, discovered in a periodicity search for the sources in the 1993 {\it ASCA} archive \citep{torii98,gotthelf98}. Further  {\it ASCA} observations in 1997 did not detect it, while a  fainter source, \srcasca, positionally consistent with the pulsar, was found in an {\it ASCA} observation of the Galactic Plane Survey carried out in 1999. An association between the two sources, which would imply a flux decrease of a factor $\sim$10, could not be confirmed because no pulsations were found in \srcasca\   \citep{vasisht00}.
 \srcasca\  is located at the center of the radio-shell supernova remnant (SNR) G29.6+0.1 \citep{gaensler99}. 
 A source consistent in flux and position with \srcasca\ was also seen with the {\it BeppoSAX} satellite in 2001 \citep{israel04short}.
 
 \begin{table}
  \begin{center}
\footnotesize
   \caption{Log of the \textit{{\it XMM-Newton}} and {\it Chandra} observations.} 
      \label{log}
   \begin{tabular}{ c c c c }
\hline 
  Satellite & Obs.ID. & Observation date & Exposure (ks)\\
\hline
\textit{{\it Chandra}} & 7578 &  2007-02-19  & 4.7  \\
\textit{{\it Chandra}} & 7579 &  2007-04-22  & 5.0  \\
\textit{{\it Chandra}} & 7580 &  2007-06-08  & 4.8  \\
\textit{{\it Chandra}} & 7581 &  2007-08-04  & 5.2  \\
\textit{{\it Chandra}} & 7582 &  2007-09-18  & 5.1  \\
\textit{{\it Chandra}} & 7583 &  2007-11-05  & 5.2  \\
{\bf \textit{{\it Chandra}}} & { 11801} &  { 2010-06-17}  & { 32}  \\
\textit{{\it XMM-Newton}} & 0602350101 &  2010-04-14  & 61  \\
\textit{{\it XMM-Newton}} & 0602350201 &  2010-04-16  & 43  \\
\hline
\end{tabular}
\end{center}
\end{table}

\begin{figure*}
\includegraphics[width=17.8cm]{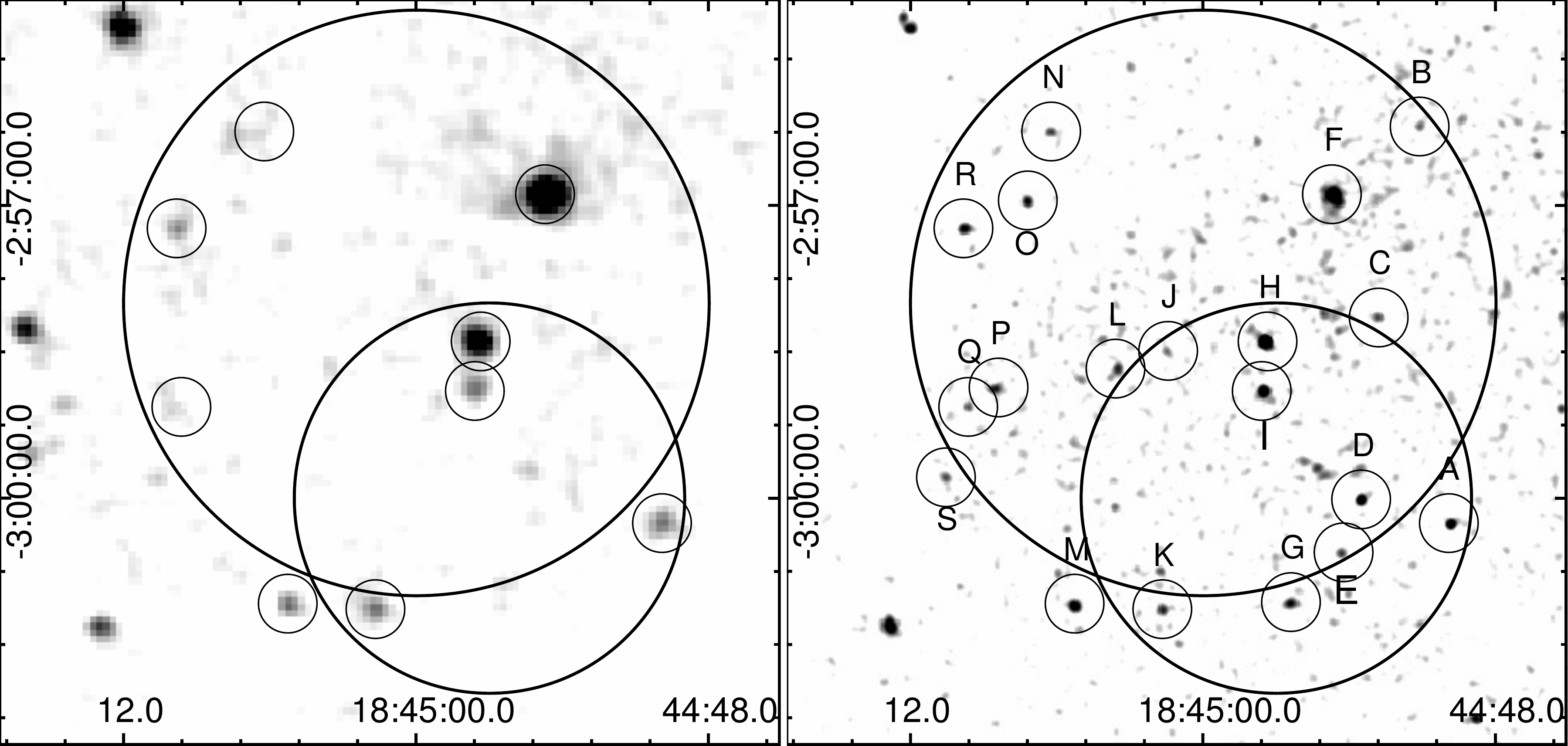}
    \caption{{\it Left: EPIC} image in the 0.3--10 keV energy range, obtained by stacking the pn and     MOS data of the two {\it XMM-Newton} observations. 
The largest circle (3$'$ radius) is the error region of \src\ reported by \citet{gotthelf98}, while the smaller one (2$'$ radius) is that reported by \citet{torii98}.
     {\it Right:} Stacking of all the 2007 and 2010 {\it Chandra} observations in the 0.3--10 keV energy range. \newline
     { In both images, the small circles (18'' for graphical purposes only) indicate the sources found in this work}.
A smoothing has been applied to both images  for display purposes.}
    \label{images}
\end{figure*}

\begin{table*}
  \begin{center}
\footnotesize
   \caption{Summary of the X-ray sources detected in the \textit{{\it XMM-Newton}} and {\it Chandra} observations.} 
\scalebox{0.85}{\begin{minipage}{24cm}
      \label{sources}
   \begin{tabular}{l l c c c c c l}
\hline 
Source & Name. & R.A. & Dec.  & Band$^b$ & Rate ({\it Chandra})$^c$ & Rate ({\it XMM/EPIC-pn})$^c$ & Other information$^d$\\
\hline
A & CXOU J184449.8--030015  & $18^{\text{h}}44^{\text{m}}49^{\text{s}}.8$ & $-3\degree00'15''.4$  & {\sc s} &$(7.1\pm1.4)\times10^{-4}$ & $(4.2\pm0.6)\times10^{-3}$ & J $=14.30$, K $=13.06$; late type star\\
B & CXOU J184451.1--025611 & $18^{\text{h}}44^{\text{m}}51^{\text{s}}.1$ & $-2\degree56'11''.5$  & {\sc s} &$(2.3\pm0.7)\times10^{-4}$&--  &  J $=8.96$, K $=8.63$; late type star\\
C & CXOU J184452.8--025809 & $18^{\text{h}}44^{\text{m}}52^{\text{s}}.8$ & $-2\degree58'09''.0$  & {\sc t} &$(2.0\pm0.7)\times10^{-4}$ & --& R1 $=17.97$, I $=16.29$; star(?) \\
D  & CXOU J184453.5--030000 & $18^{\text{h}}44^{\text{m}}53^{\text{s}}.5$ & $-3\degree00'00''.8$  & {\sc t} &$(6.6\pm1.3)\times10^{-4}$ &-- &  J $=12.96$, K $=11.63$; late type star\\
E & CXOU J184454.3--030033 & $18^{\text{h}}44^{\text{m}}54^{\text{s}}.3$ & $-3\degree00'33''.3$  & {\sc t} &$(4.3\pm1.2)\times10^{-4}$&-- & J $=15.31$, K $=14.31$; late type star \\
F & \srcasca$^a$ & $18^{\text{h}}44^{\text{m}}54^{\text{s}}.7$ & $-2\degree56'53''.2$ & {\sc h} &$(4.8\pm0.3)\times10^{-3}$  &$(2.34\pm0.09)\times10^{-2}$  &   \\
G& CXOU J184456.4--030104 & $18^{\text{h}}44^{\text{m}}56^{\text{s}}.4$ & $-3\degree01'04''.6$ & {\sc s} &$(4.0\pm1.0)\times10^{-4}$&-- & J $=13.41$, K $=13.07$; star(?)\\
H& CXOU J184457.5--025823 & $18^{\text{h}}44^{\text{m}}57^{\text{s}}.5$ & $-2\degree58'23''.7$ & {\sc t} & $(2.2\pm0.2)\times10^{-3}$& $(1.1\pm0.1)\times10^{-2}$ &  \\
I & CXOU J184457.6--025854 & $18^{\text{h}}44^{\text{m}}57^{\text{s}}.6$ & $-2\degree58'54''.6$   & {\sc s} &$(7.3\pm1.2)\times10^{-4}$ & $(1.9\pm0.4)\times10^{-3}$ & J $=10.27$, K $=9.31$; late type star \\
J  & CXOU J184501.5--025829 & $18^{\text{h}}45^{\text{m}}01^{\text{s}}.5$ & $-2\degree58'29''.5$   & {\sc s} & $(1.5\pm0.6)\times10^{-4}$ & -- &  \\
K& CXOU J184501.7--030108 & $18^{\text{h}}45^{\text{m}}01^{\text{s}}.7$ & $-3\degree01'08''.2$  & {\sc t} &$(3.9\pm1.1)\times10^{-4}$& $(3.1\pm0.5)\times10^{-3}$ & J $=14.75$, K $=11.95$; late type star  \\
L & CXOU J184503.8--025845 & $18^{\text{h}}45^{\text{m}}03^{\text{s}}.8$ & $-2\degree58'45''.6$  & {\sc s} &$(1.6\pm0.6)\times10^{-4}$ &-- &   J $=13.01$, K $=12.28$; late type star\\
M& CXOU J184505.3--030105 & $18^{\text{h}}45^{\text{m}}05^{\text{s}}.3$ & $-3\degree01'05''.4$ & {\sc t} &$(7.9\pm1.5)\times10^{-4}$& $(3.6\pm0.7)\times10^{-3}$ & J $=13.77$, K $=13.08$; late type star  \\
N& CXOU J184506.3--025614 & $18^{\text{h}}45^{\text{m}}06^{\text{s}}.3$ & $-2\degree56'14''.6$  & {\sc t} &$(3.7\pm1.3)\times10^{-4}$ & $(2.0\pm0.4)\times10^{-3}$& J $=15.29$, K $=12.80$; late type star  \\
O& \srccc$^a$ & $18^{\text{h}}45^{\text{m}}07^{\text{s}}.2$ & $-2\degree56'57''.4$  & {\sc s} & $(6.4\pm1.6)\times10^{-4}$ & -- & J $=13.73$, K $=12.71$; late type star \\
P  & CXOU J184508.5--025852 & $18^{\text{h}}45^{\text{m}}08^{\text{s}}.5$ & $-2\degree58'52''.0$   & {\sc t} & $(5.1\pm1.3)\times10^{-4}$ &--  & J $=16.01$, K $=11.72$; late type star \\
Q& CXOU J184509.7--025903 & $18^{\text{h}}45^{\text{m}}09^{\text{s}}.7$ & $-2\degree59'03''.9$  & {\sc h} & -- & $(1.8\pm0.4)\times10^{-3}$ &   \\
R& \srcccc$^a$ & $18^{\text{h}}45^{\text{m}}09^{\text{s}}.8$ & $-2\degree57'14''.1$   & {\sc h} & $(8.5\pm2.1)\times10^{-4}$ & $(1.2\pm0.4)\times10^{-3}$ &  \\
S  & CXOU J184510.6--025948 & $18^{\text{h}}45^{\text{m}}10^{\text{s}}.6$ & $-2\degree59'48''.3$   & {\sc s} & $(2.4\pm0.8)\times10^{-4}$ &  --&  J $=15.79$, K $=13.39$; late type star\\
\hline
\end{tabular}
\end{minipage}}
\end{center}

\begin{flushleft} $^a$ Also reported in \citet{tam06}.  \\
$^b$ Detection band: {\sc t}: 0.3--10 keV; {\sc s}: 0.3--2 keV; {\sc h}: 2--10 keV.  \\
$^c$ Count rate in the given detection band.\\
$^d$ Magnitude in J and K infrared bands from the 2MASS catalogue, or R1 and I optical bands from the USNO B1 catalogue.
\end{flushleft}
\end{table*}

Based on the value of its spin period, on the  soft X-ray spectrum (a blackbody with temperature kT$\sim$0.6 keV or a steep power law with photon index $\Gamma\sim$5), and on  the possible association with a     SNR (if indeed     \srcasca\ and the pulsar are the same source), it was suggested that \src\ could belong to the class of  anomalous X-ray pulsars. These sources, together with the soft gamma-ray repeaters which show similar properties, are generally believed to be isolated neutron stars powered by strong magnetic fields, i.e. magnetars \citep[see e.g.][]{mereghetti08}.    

\citet{tam06} analyzed 7 {\it Chandra} observations taken between June and August 2003, and found  three  X-ray sources inside the large  (3$'$ radius) error box of \src\ reported in \citet{gotthelf98}, the brightest  one coinciding with \srcasca . 
Its high   absorption  (N$_\text{H}>10^{22}$ cm$^{-2}$) was consistent with that of the pulsar seen in 1993, but again no pulsations could be detected. The other two {\it Chandra} sources 
were   too faint for a detailed spectral and timing analysis. 

These results indicate that \src\ might be a transient magnetar, which experienced an outburst shortly before the 1993 {\it ASCA} observation  and subsequently faded  to quiescence. 
Similar behaviours are not unusual in magnetars \citep[see, e.g.][]{rea11}.

Here we present a new investigation of all the X-ray sources in the sky region of  \src , based on {\it XMM-Newton}  and  {\it Chandra}  archival data. These observations allowed us to carry out a more detailed spectral and timing analysis of the sources already reported in the literature and to discover other possible counterparts of \src .

\section{Observations and data reduction}
\label{data_reduction}

We used two {\it XMM-Newton} observations, with durations  of 61 and 43 ks, carried out in April 2010 (see details in Table~\ref{log}). The three cameras of the EPIC instrument (one pn camera, \citealt{struder01short}) and two   MOS  cameras \citep{turner01short}  were operated in full-frame mode in both observations. { The corresponding EPIC-pn and MOS read-out time resolution is 0.073 and 2.6 s, respectively}.
The data were reduced with SAS v. 14.0.0. We selected single- and double-pixel events ({\sc pattern}$\leq$4) for the pn and single-  and multiple-pixel events for the MOS ({\sc pattern}$\leq$12). Time intervals with high particle background were excluded from the analysis, resulting in a net exposure time of $\sim$40 and $\sim$32 ks for the first and the second observation, respectively. 
{ We excluded from this work the 2003 {\it XMM-Newton} observation (Obs.ID: 0046540201) because is affected by high particle background which limits the net exposure time to a few ks.}

The { seven} {\it Chandra} observations used in this work were performed in 2007 and { 2010}, for a  total exposure time of {$\sim$60 ks} (see Table~\ref{log}). They were made using the ACIS-S detector in full-frame mode,  yielding a time resolution of 3.241 s. 
We reduced the data  with the {\sc ciao} software v.4.7 and the {\sc caldb} v.4.6.9. 

{ For {\it XMM-Newton} data, we  extracted the source counts  from circular regions with radius 20$''$ (except in a few cases mentioned below) and the background counts from nearby  source-free circular regions with 40$''$ radius. For {\it Chandra} data, we instead used circular regions with radius of 5$''$ and 20$''$ for source and background, respectively.}
Spectral fits were carried out with    {\sc xspec} v.12.8.2 in the energy range 0.3--10 keV. In the following,  all the errors on the spectral parameters are at the 90\% c.l. . For the timing analysis, the times of arrival of the counts were converted to the Solar System barycenter using  the JPL planetary ephemerides DE405 and the {\it Chandra} coordinates of the sources.

\section{Data analysis and results}

We first created a combined EPIC image in the energy range 0.3--10 keV by stacking the pn and MOS data of both observations (Fig.~\ref{images}-{\it left}).
In addition to the three  {\it Chandra} sources reported in \citet{tam06} (labelled here as source F, O and R),  the EPIC data reveal the presence of 7 new sources inside (or slightly outside) the error regions of \src. All of them are detected at $\geq3\sigma$ in at least one of the considered  energy ranges (0.3--10 keV,  0.3--2 keV, and 2--10 keV).  
The brightest of the new sources ({\it H} and {\it I}) are positionally coincident with the region of diffuse X-ray emission detected by \citet{vasisht00} in the {\it ASCA}-SIS data, which had a spatial resolution insufficient  to resolve them.

We created a combined image in the 0.3--10 keV energy range by joining the data of  the seven  {\it Chandra} observations (Fig.~\ref{images}-{\it right}). This image shows all the sources detected by {\it XMM-Newton} { (except for source Q)} plus 10 fainter ones.  Note that the 2003 {\it Chandra} observations covered with high sensitivity only a small fraction of the pulsar error region; this explains why most of these sources were not reported by \citet{tam06}.

 We also looked in the optical USNO B1.0 \citep{monet03short} and infrared (IR) 2MASS \citep{skrutskie06short} catalogues for any possible counterpart of all the detected sources. The coordinates, count rates and possible counterparts of the sources are listed in Table~\ref{sources}.

\begin{table*}
  \begin{center}
     \caption{EPIC spectral results. Errors are at $90\%$ for each parameter of interest.} 
      \label{spectral_model}
\begin{tabular}{lclccccc}
\hline
Name &Source & Model & N$_\text{H}$&  $\Gamma$& kT$_{\text{bb}}$/kT$_{\text{apec}}$  & Flux$^a$ & $\chi^2$/dof \\
&&  & ($10^{22}$ cm$^{-2}$) &  & (keV) & (10$^{-13}$ erg cm$^2$ s$^{-1}$) &  \\
\hline

\srcasca & \multirow{2}{*}{F} & {\sc powerlaw} & $9.7^{+2}_{-1.7}$ & $1.2\pm0.3$ & -- & $3.1\pm0.2$  & 23.10/16 (0.11$^b$) \\
 && {\sc bbody} & $5.8^{+1.3}_{-1.1}$ & -- & $2.1^{+0.3}_{-0.2}$ & $2.9\pm0.2$  & 27.12/16 (0.04$^b$) \\

\hline
CXOU J184457.5--025823 & \multirow{2}{*}{H} & {\sc powerlaw} & $0.8^{+0.4}_{-0.3}$ & $1.3\pm0.3$ & -- & $0.57\pm0.07$  & 8.06/9 \\
 & & {\sc bbody} & $0.14^{+0.16}_{-0.12}$ & -- & $1.2^{+0.1}_{-0.1}$ & $0.45^{+0.05}_{-0.06}$  & 8.95/9 \\

\hline
CXOU J184457.6--025854 & \multirow{3}{*}{I} & {\sc powerlaw} & $0.72^{+0.32}_{-0.65}$ & $4.8^{+1.7}_{-3.8}$ & -- & $0.041_{-0.012}^{+0.018}$  & 17.24/11 \\
 & & {\sc bbody} & $0.2^{+0.2}_{-0.5}$ & -- & $0.26^{+0.12}_{-0.13}$ & $0.036^{+0.009}_{-0.011}$  & 18.81/11 \\
 & & {\sc apec} & $0.75^{+0.19}_{-0.29}$ & -- & $0.8\pm0.2$ & $0.042\pm0.008$  & 6.08/11 \\
 \hline

\srcccc & \multirow{2}{*}{R} & {\sc powerlaw} & \multirow{2}{*}{1.69$^c$} & $0.9_{-0.1}^{+1.5}$ & -- & $0.22_{-0.08}^{+0.09}$  & 1.14/2 \\
 & & {\sc bbody} & & -- & $1.5^{+1.0}_{-0.4}$ & $0.18^{+0.09}_{-0.06}$  & 1.04/2 \\

\hline

\end{tabular}
\end{center}
\begin{flushleft}$^a$ Absorbed flux in the 0.3--10 keV energy band.\\
$^b$ Null hypothesis probability.\\
$^c$ Fixed.
 \end{flushleft}
\end{table*}

\subsection{\srcasca\ (Source F)}
\label{spectral_analysis_1}

{ \srcasca\ is the brightest source presented in \citet{tam06}}. For the spectral analysis of this source, we first considered the two {\it XMM-Newton} observations separately. By fitting simultaneously the  pn and MOS spectra with simple models  (powerlaw, blackbody)  we obtained  acceptable fits and found no evidence for time variations in the flux or spectral parameters.  Therefore, also considering the short time interval between the two observations, we joined them and extracted a single spectrum combining pn+MOS data using the  SAS tool {\sc epicspeccombine}. { This allowed us to collect a total of $\sim$2300 net counts in the 0.3--10 keV energy band.}

 \begin{figure}
	\subfigure{\includegraphics[height=8.7cm,angle=270,]{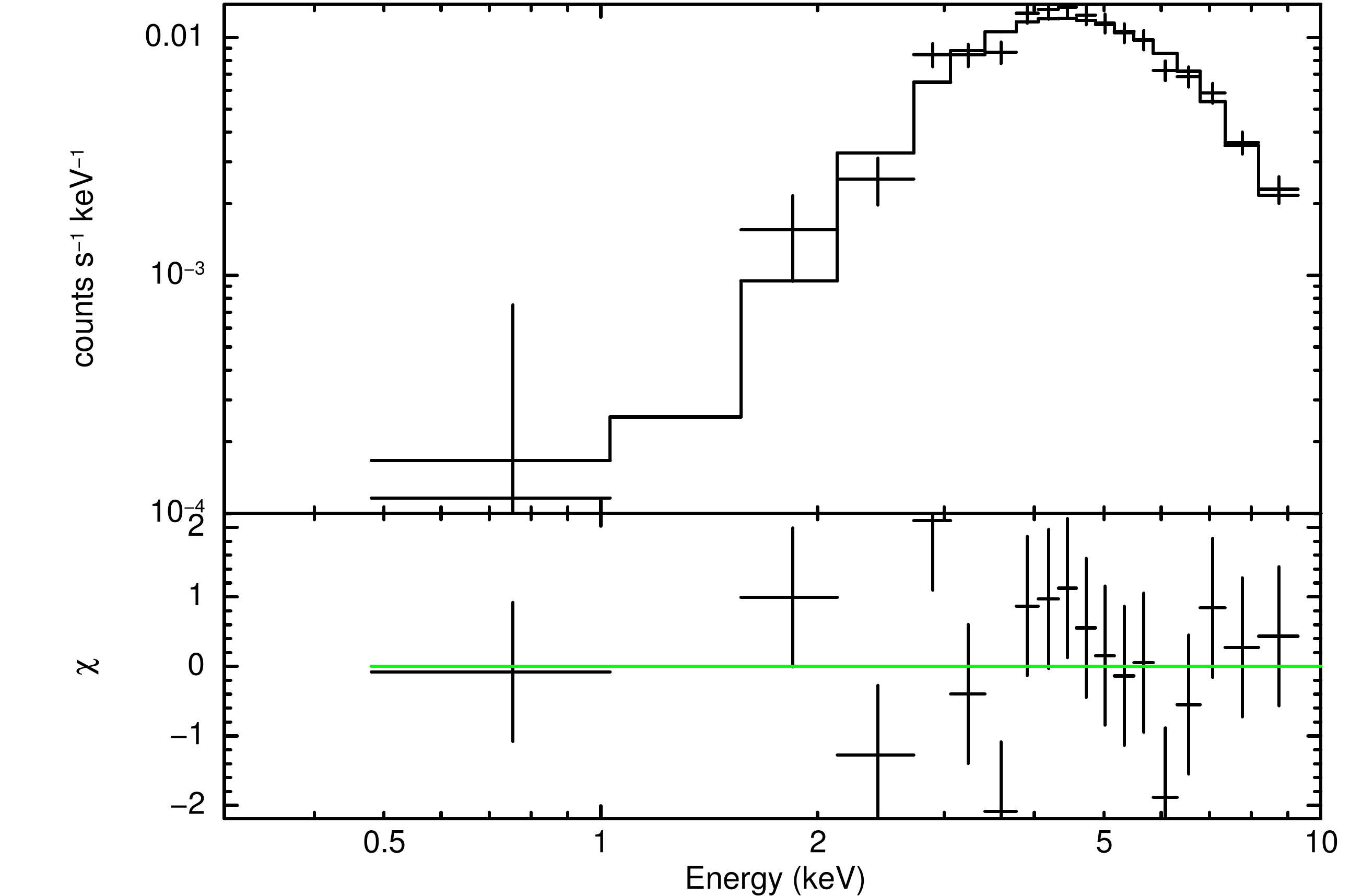}}
    \caption{EPIC spectrum of \srcasca\ (source {\it F}). Top panel: data and  best-fit powerlaw model. Bottom panel: residuals in units of $\sigma$.}
        \label{pow1}
\end{figure}

In  the individual observations, both a power law (photon index $\Gamma\sim1$)  and  a blackbody  (temperature in the range $2.1-2.4$ keV) gave equally  acceptable fits. However, the fit to the combined spectrum rebinned with a minimum of 200 counts per bin favours the power-law model (see Figure~\ref{pow1} and Table~\ref{spectral_model}). 

The source count rates measured in the single {\it Chandra} observations were consistent with a constant, indicating no significant flux variability. Hence, for the {\it Chandra} spectral analysis, we created a stacked  spectrum using all the  observations and   fitted it with a powerlaw, fixing N$_\text{H}$ to the value found with {\it XMM-Newton}.  
{ The resulting photon index ($0.9\pm0.5$) and   absorbed flux ($(3.3\pm0.75)\times10^{-13}$ erg cm$^{-2}$ s$^{-1}$, 2--10 keV) are consistent with those found with {\it XMM-Newton}. }

\begin{figure}
	\subfigure{\includegraphics[angle=270,width=8.7cm]{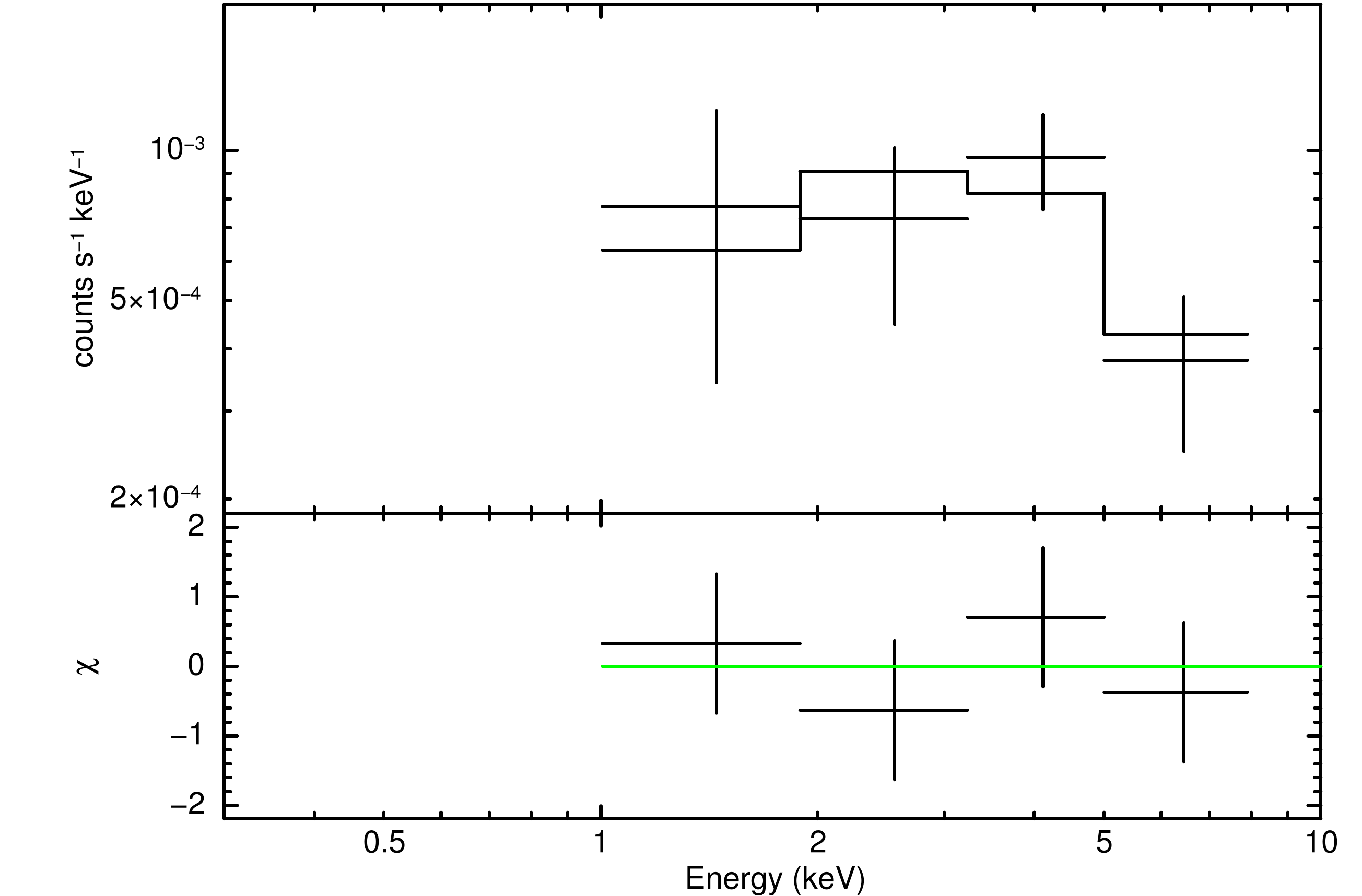}}
    \caption{EPIC spectrum of source {\it R} stacking the spectra of the first and second {\it XMM-Newton} observations. Top panel: data and  the best-fit powerlaw model. Bottom panel: residuals in units of $\sigma$.}
    \label{powerlaw_c}
\end{figure}

In order to search for pulsations in \srcasca, we used the EPIC pn and MOS source counts of the two observations. We used only the counts with energy above 2  keV, where the the signal to noise ratio is the highest. This yielded a total of about 1700 counts, of which $\sim$16\% can be attributed to the background. Using a Rayleigh test technique, we explored periods in the range 6.5--7.5 s, which accounts for a possible spin-up or spin-down of the source of $|\dot P| < 10^{-9}$ s s$^{-1}$ from 1993 up to the present days.  No significant pulsations were detected and, by means of Montecarlo simulations assuming a sinusoidal pulse profile, we could set a 3$\sigma$ c.l. upper limit of 6\% on the source pulsed fraction (defined as the amplitude of the sinusoid divided by its average value). 

\begin{figure*}
	\subfigure{\includegraphics[angle=270,width=8.8cm]{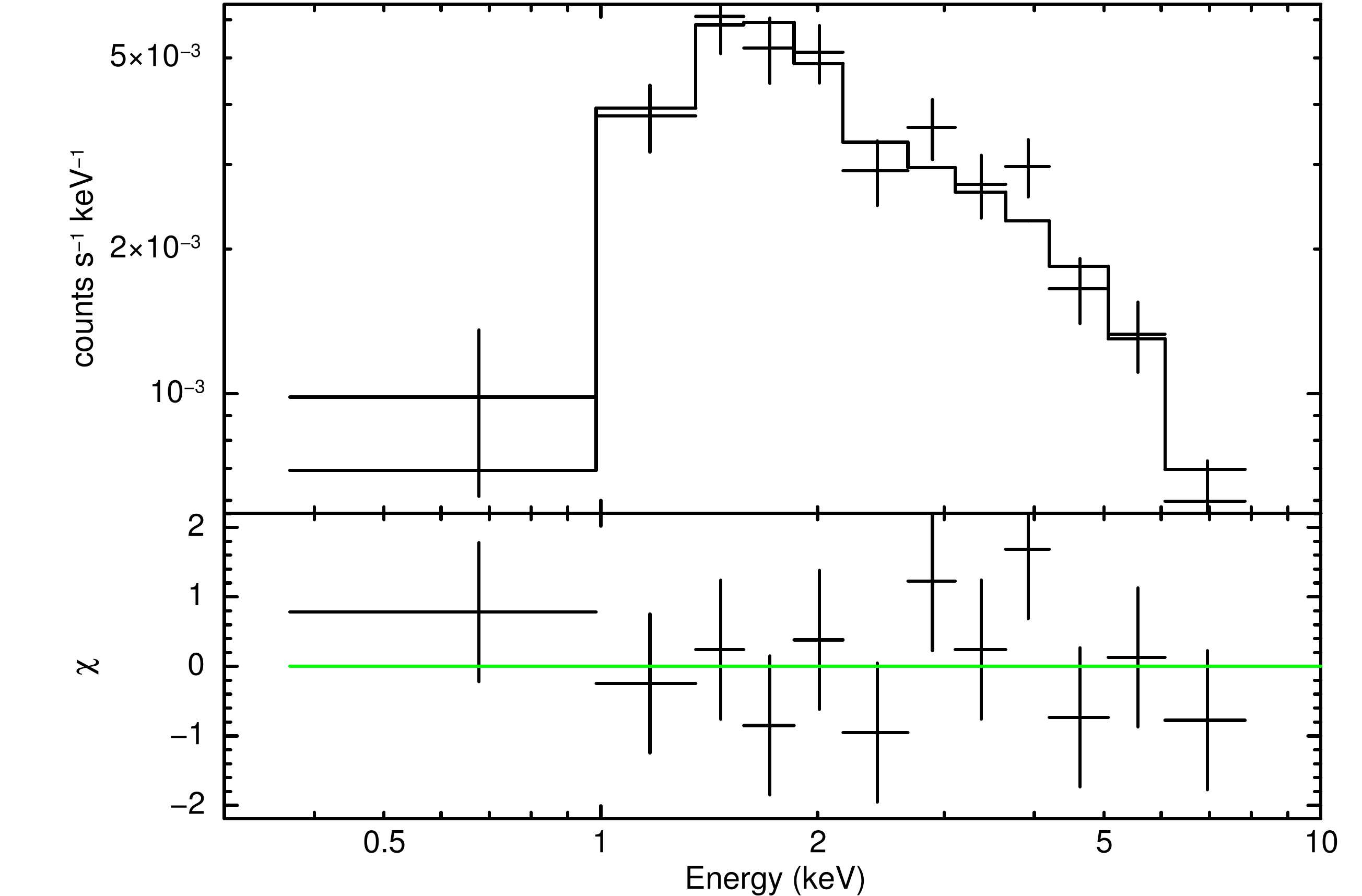}}
	\subfigure{\includegraphics[angle=270,width=8.8cm]{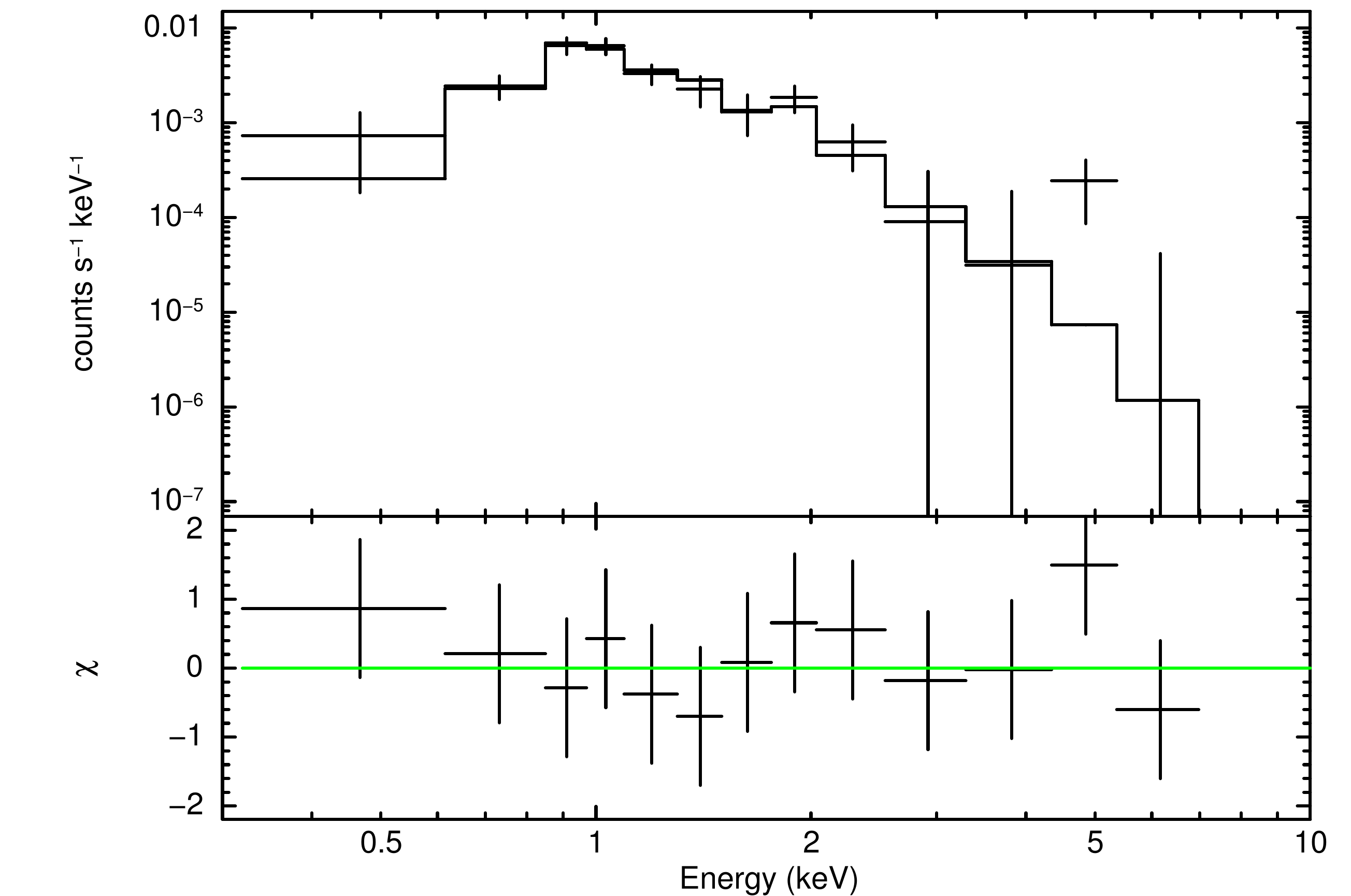}}
    \caption{EPIC spectrum of of source {\it H}-({\it left}) and {\it I}-({\it right}) stacking the spectra of the first and second {\it XMM-Newton} observations. Top panels: data and  the best-fit powerlaw (left) and {\sc apec} (right) models. Bottom panels: residuals in units of $\sigma$.}
    \label{apec3}
\end{figure*}

\subsection{\srccc\  (Source O) }
\label{spectral_analysis_4b}
 
This  source was first reported by \citet{tam06}, who noted its positional  coincidence with a near infrared object with magnitude K=12.7      of  the 2MASS catalogue. In {\it XMM-Newton}, we could not study it in detail since it fell very close to the gap between CCDs in the EPIC pn data. In the {\it Chandra} observations, the source is detected only in the soft energy band (0.3--2 keV). This finding and the association with a bright IR object suggest that this source is most likely a late type star. 

\subsection{\srcccc\ (Source R) }
\label{spectral_analysis_4}

This was the faintest of the three  {\it Chandra} sources previously reported by  \citet{tam06}.
Due to its faintness   we could perform only a rough  spectral analysis using the stacked EPIC spectrum of the two {\it XMM-Newton} observations { yielding about 180 spectral net counts in the 0.3--10 keV energy band.} The source is hard and most of its photons are above 1.5 keV, making the estimates of the absorption poorly constrained. Therefore, we fixed the absorption at the   value N$_\text{H}$ = $1.69\times10^{22}$ cm$^{-2}$, corresponding to the total column density  in the direction of the source  \citep{dickey90}. We then fitted the spectrum with an absorbed powerlaw or a blackbody model. We found statistically acceptable results with both models (Table~\ref{spectral_model}).
Adopting the powerlaw model (Figure~\ref{powerlaw_c}), we estimated a source absorbed flux of $(2.2\pm0.9)\times 10^{-14}$ erg cm$^{-2}$ s$^{-1}$ in the energy range 0.3--10 keV.
The   spectrum of this source appears harder than expected for a magnetar in quiescence.

\subsection{CXOU J184457.5--025823 (Source H)}
\label{spectral_analysis_2}

This source, reported here for the first time,  is the second brightest object inside the error box of \src , only a factor of $\sim$2--3 fainter than \srcasca .  To avoid contamination from   source I, we used an extraction radius of  15$''$. 
Since no evidence of variability was seen by comparing the spectra of the two {\it XMM-Newton} observations, we summed the pn and MOS spectra of the two observations, { obtaining a total of 755 net counts in the 0.3--10 keV energy band.}

   The resulting spectrum, rebinned with a minimum of 100 counts per bin,  was well fit by  either a  blackbody  with   kT $=1.2\pm0.1$ keV or by a  power law with  $\Gamma=1.3\pm0.3$ (Table~\ref{spectral_model}), while a thermal plasma model ({\sc apec} in {\sc xspec}) gave a bad fit. 
From the powerlaw best-fit (Fig.~\ref{apec3}-{\it left}), we measured an absorbed flux  of $(5.7\pm0.7) \times10^{-14}$ erg cm$^{-2}$ s$^{-1}$, in the 0.3--10 keV energy range.

Due to   the limited counting statistics of the  {\it Chandra} data, we extracted a spectrum by summing the counts of all the observations and fitted it with a power-law keeping  the  column density fixed at the value N$_{\text{H}} = 8\times10^{21}$ cm$^{-2}$ derived with {\it XMM-Newton}. 
{ This yielded a photon index  $\Gamma=1.1\pm0.8$ and an absorbed  flux of  $(7.4\pm5) \times10^{-14}$ erg cm$^{-2}$ s$^{-1}$ in the  0.3--10 keV range, implying  that the source remained quite stable between 2007 and 2010.}

We carried out a search for pulsations in source H, with the same procedure described above for \srcasca\ but, in this case, using the counts in the 1--12 keV energy range. Again no significant signals were found and we derived a 3$\sigma$ c.l. upper limit of 18\% on the source pulsed fraction for periods in the range 6.5--7.5 s. 

\begin{figure}
\subfigure{\includegraphics[width=8.5cm]{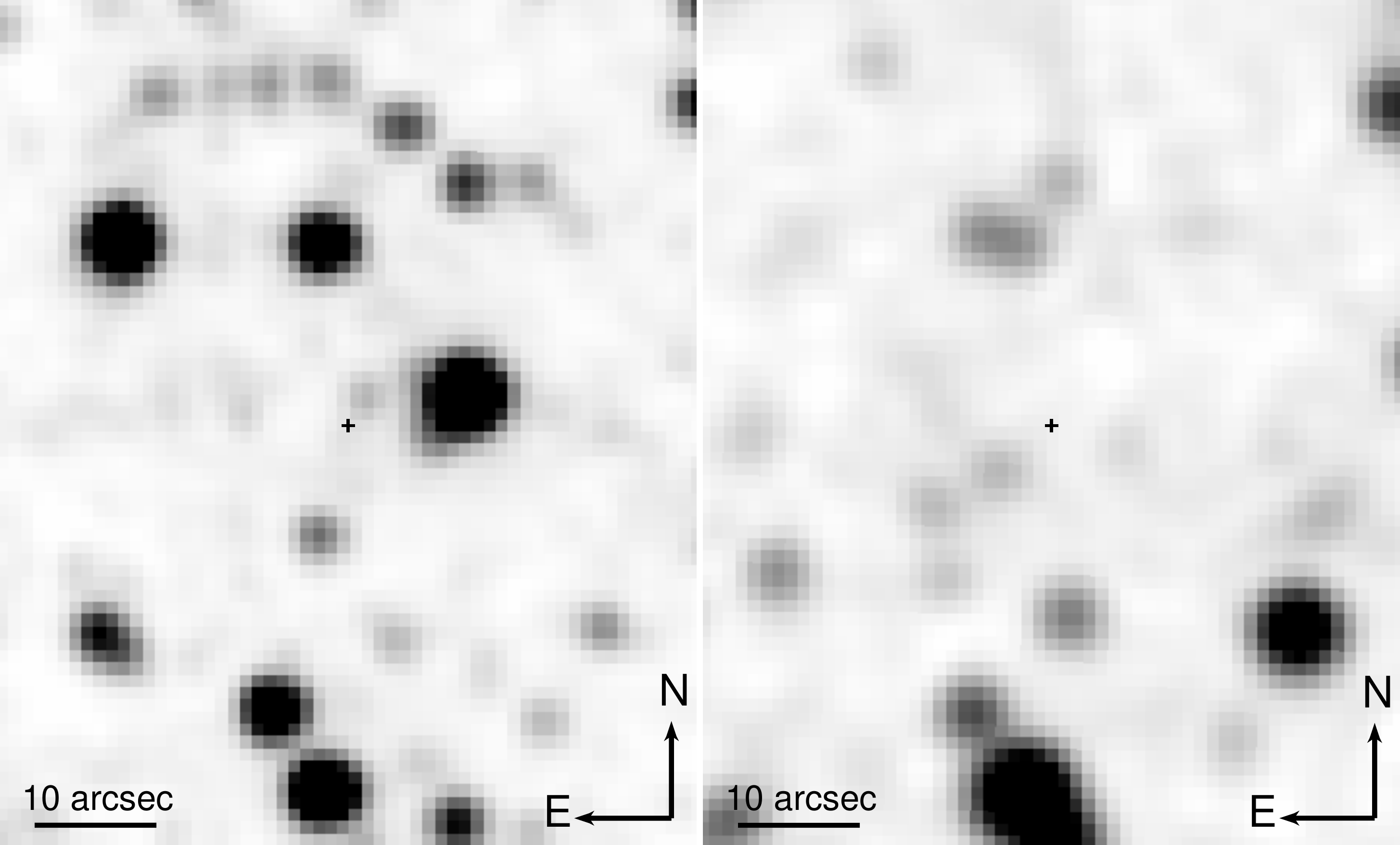}}
    \caption{2MASS K-band image (left) and DSS blue band (right) image of the field around source {\it H}. The cross is the source position with an uncertainty of 1$''$. }
    \label{2mass_D}
\end{figure}

Finally, we remark that we could not associate any optical or infrared counterpart to the source (see Figure~\ref{2mass_D}).

\subsection{CXOU J184457.6--025854 (Source I)}
\label{spectral_analysis_3}

This source was too  faint for a spectral analysis of the  individual  {\it XMM-Newton} observations. 
We extracted, from a region of 15$''$  radius, a total pn+MOS source spectrum by summing the two {\it XMM-Newton} observations { which gave us a total of 220 net counts in the 0.3--10 keV energy band.}
The fits with either a powerlaw or a blackbody were acceptable but left several residuals ($\chi^2_{\nu}>1.5$), while a good fit could be obtained with the  {\sc apec}  model (Figure~\ref{apec3}-{\it right} and Tab.~\ref{spectral_model}). This gave a plasma temperature of 0.8 keV  and  an absorbed 0.3--10 keV flux of $(4.2 \pm 0.8) \times10^{-15}$ erg cm$^{-2}$ s$^{-1}$. 
This  source was too faint for a spectral analysis with the  {\it Chandra} data.

Source {\it I} can be associated to an IR counterpart in 2MASS with  magnitudes J=$10.267\pm0.023$, H=$9.542\pm0.025$ and K=$9.305\pm0.025$.  An optical counterpart can also be found with a B1-band magnitude of $15.29\pm1$. Based on the optical/IR colours and the thermal X-ray spectrum, we conclude that this object is most likely a foreground star of spectral type K or M.

\subsection{CXOU J184505.3--030105 (Source M)}

Source {\it M}, positionally coincident with a 2MASS object with magnitudes of J=13.8, H=12.6 and K=13.1, can be likely associated to a foreground late type star. Its combined (pn+MOS of both observations) X-ray spectrum (218 net counts in the 0.3--10 keV band) is poorly constrained but can be well modelled ($\chi^2_{\nu}<1$) with an {\sc apec} component. The best fit parameters are N$_\text{H}=(5.7_{-3}^{+6})\times10^{22}$ cm$^{-2}$, kT$=1.9_{-1}^{+4}$ keV and a 0.3--10 keV absorbed flux of $(1.6\pm0.5) \times 10^{-14}$ erg cm$^{-2}$ s$^{-1}$

\subsection{CXOU J184501.7--030108 (Source K)}

Also source {\it K} can be associated to an IR source of the 2MASS catalogue with magnitudes in J, H and K bands of $14.745\pm0.057$, $12.779\pm0.042$ and $11.947\pm0.035$, respectively. No counterparts are reported in optical catalogues. The two {\it XMM-Newton} observations yielded $\sim$300 net counts and its X-ray spectrum can be well described with a single blackbody ($\chi^2/dof=7.09/7$) while a powerlaw or an {\sc apec} model are statistically worse. We found N$_\text{H}<1.5\times10^{21}$ cm$^{-2}$, kT$=0.9_{-0.1}^{+0.2}$ keV and a 0.3--10 keV absorbed flux of $(1.3\pm0.4) \times 10^{-14}$ erg cm$^{-2}$ s$^{-1}$. Also in this case, a possible association to a foreground late type star is likely.

\subsection{CXOU J184449.8--030015 (Source A)}

Source {\it A} has an IR counterpart in the 2MASS catalogue with magnitudes in J, H and K bands of $14.296\pm0.037$, $13.472\pm0.037$ and $13.06\pm1$, respectively. An optical counterpart is also associated to the X-ray source, with magnitudes in R1, B1 and I bands of $14.6\pm1$, $19.6\pm1$ and $15.25\pm1$, respectively. The combined source X-ray spectrum ($\sim$200 total net counts) could be well fitted with a blackbody or an {\sc apec} model. In the latter case, we found a N$_\text{H}=(8_{-6}^{+12})\times10^{21}$ cm$^{-2}$, kT$=3.8_{-2.3}^{+3.7}$ keV and a 0.3--10 keV absorbed flux of $(1.2\pm0.6) \times 10^{-14}$ erg cm$^{-2}$ s$^{-1}$. Because of its spectral properties and IR/optical emission, the source may be associated to a late type star.

\section{Discussion and Conclusions}

\begin{figure}
\subfigure{\includegraphics[width=8.8cm]{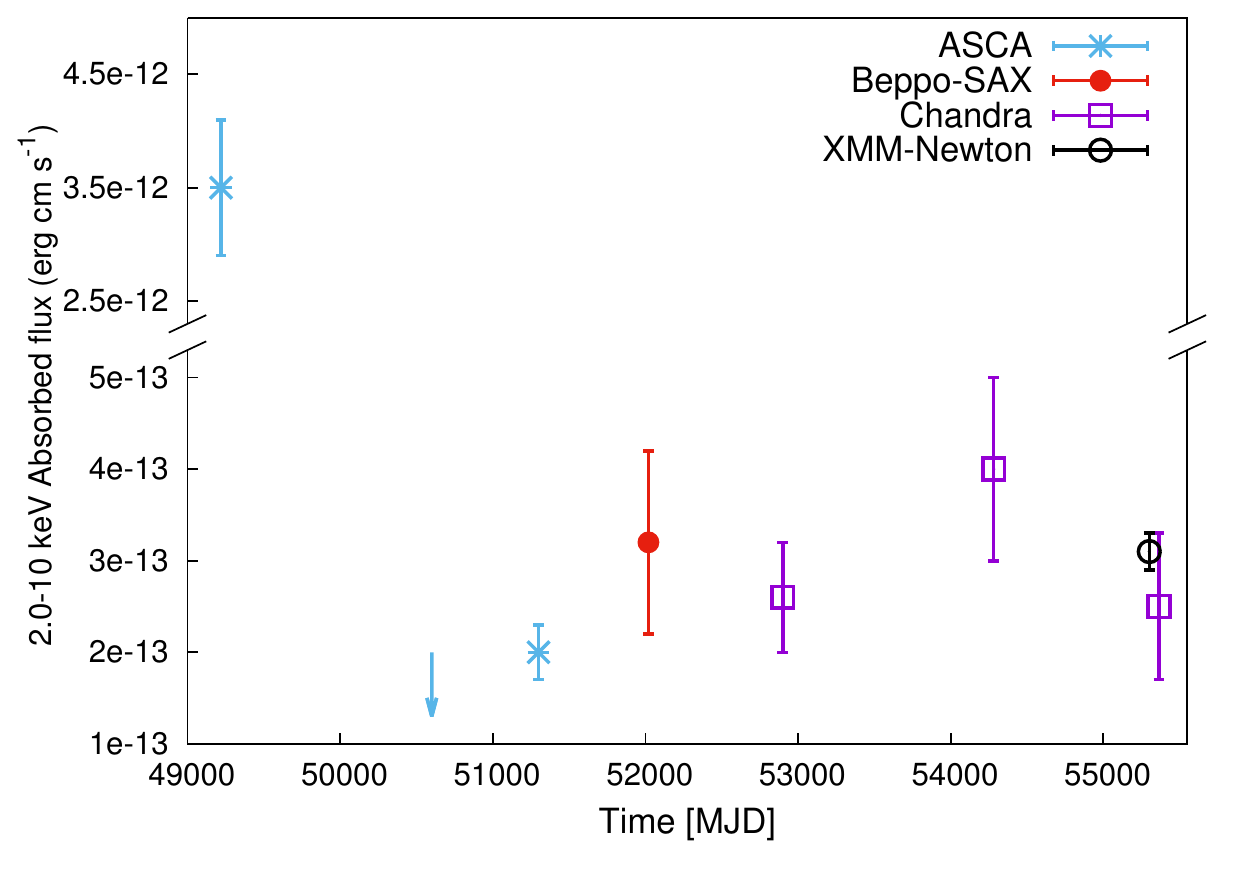}}
    \caption{Absorbed 2--10 keV light curve of \srcasca. The {\it ASCA}, {\it BeppoSAX} and 2003 {\it Chandra} fluxes are taken from \citet{tam06}. { The {\it BeppoSAX} and {\it ASCA} fluxes might be overestimated due to the presence of other unresolved sources.}}
    \label{lc_tot}
\end{figure}

After more than 20 years since its discovery, the nature of the pulsating X-ray source \src\ is still unknown: it could be either a transient magnetar or a transient accreting X-ray binary  containing a neutron star or a white dwarf.
All the X-ray observations carried out after the 1993 discovery revealed only fainter sources, implying a variability of at least one order of magnitude, but none of  them could be safely associated  with \src\ since the  pulsations at 7s were never  detected again.  

The brightest of these sources, \srcasca , remains a good candidate for being the pulsar counterpart. The location at the center of the supernova remnant G29.6$+$0.1, the lack of a bright optical/IR counterpart,  and the  long term light curve  (see Fig.~\ref{lc_tot}) strongly support the interpretation in terms of  a transient  magnetar.  Our new upper limit on the pulsed fraction in \srcasca\   is incompatible with the strong modulation observed in \src . However, magnetars often show changes in their pulse profile when they evolve from an outburst toward quiescence, and some of them have  pulsed fractions comparable with our upper limits (for example SGR 1806--20; e.g. \citealt{mereghetti05}c,\citealt{woods07}). 
The nearly constant flux that the source has maintained for almost 16--17 years ($\sim2.5\times10^{-13}$ erg cm$^{-2}$ s$^{-1}$, Fig.~\ref{lc_tot})  corresponds to an average luminosity of  $3\times10^{33}$ (d/10 kpc)$^2$ erg s$^{-1}$, which is fully consistent with that of quiescent magnetars, while the relatively hard spectrum is the only characteristic of \srcasca\ that disfavours this interpretation. In fact, this spectrum is harder than that typically observed in magnetars, especially when they are in a quiescent state, and  it is also harder than that measured for \src\ in 1993. 
 
Of course, it is also possible that \srcasca\ is a source totally unrelated to the pulsar. Its presence could have been easily missed in the 1993 {\it ASCA} data dominated by the much brighter X-ray  pulsar. 
Besides \srcasca , the {\it Chandra} and {\it XMM-Newton} data reported here show the presence of several faint X-ray sources. Many of them, with soft X-ray spectra and optical/IR counterparts, may be associated to foreground stars and are most likely not associated to \src. Only two of these sources were previously reported in the literature (\srccc\ and \srcccc; \citealt{tam06}). Amongst the other newly  reported objects, we note that the sources H, J, and Q are not associated to any optical/IR counterpart. 
 
Source H (CXOU J184457.5--025823), being the second brightest in the error region after \srcasca, is particularly interesting.  The lack of a bright optical/IR counterpart,  together with a  hard and  highly absorbed spectrum, indicate that this source is unlikely to be a normal field star. On the other hand, its absorption is smaller than that of \srcasca\ (and of the pulsar seen in 1993) suggesting that source H is not a background AGN.   Associating this source with the pulsar implies a variability of  at least a factor of 60, still fully compatible with the variability seen in transient magnetars, but we note that  source H lies outside the SNR. 

We finally note that, although none of the X-ray sources reported here has optical/IR counterparts compatible with a high mass X-ray binary, the possibility that \src\ is  a  neutron star  or white dwarf accreting from a low mass companion can not be excluded.

\section*{Acknowledgements}

This work has been partially supported through financial contribution from the agreement ASI/INAF/I/037/12/0 and PRIN INAF 2014.

The results are in part based on observations obtained with XMM-Newton, an ESA science mission with instruments and contributions directly funded by ESA Member States and NASA, and on data obtained from the Chandra Data Archive. 
This publication makes partially use of data products from the Two Micron All Sky Survey, which is a joint project of the University of Massachusetts and the Infrared Processing and Analysis Center/California Institute of Technology, funded by the National Aeronautics and Space Administration and the National Science Foundation.

\addcontentsline{toc}{section}{Bibliography}
\bibliographystyle{mn2e}
\bibliography{biblio}

\bsp
\label{lastpage}
\end{document}